\documentclass[dvips]{acta}
\usepackage{lscape,epsfig}
\usepackage{amssymb}
\usepackage{amsmath}
\usepackage{epstopdf}
\usepackage{graphicx}

\SetPages{1}{1}

\SetVol{68}{2018}

\begin{document}

\begin{Titlepage}

\Title{New Galactic multi-mode Cepheids from the ASAS-SN Survey}
\Author{J.~~J~u~r~c~s~i~k$^1$,~~G.~~H~a~j~d~u$^{2,3,4}$,~~and~~M.~~C~a~t~e~l~a~n$^{2,4}$\footnote{On sabbatical leave at the European Southern Observatory, Av Alonso de C\'ordova 3107, 7630355 Vitacura, Santiago, Chile}}
{$^1$ Konkoly Observatory, H-1525 Budapest PO Box 67, Hungary\\
$^2$ Instituto de Astrof\'{i}sica, Pontificia Universidad Cat\'olica de Chile, Av. Vicu\~na Mackenna 4860, 782-0436 Macul, Santiago, Chile\\
$^3$ Astronomisches Rechen-Institut, Zentrum f\"ur Astronomie der Universit\"at Heidelberg, M\"onchhofstr. 12-14, D-69120 Heidelberg, Germany\\
$^4$ Instituto Milenio de Astrof\'{i}sica, Santiago, Chile\\
e--mail: jurcsik.johanna@gmail.com; ghajdu@astro.puc.cl
}
\Received{~} 
\end{Titlepage}


\Abstract{A systematic search for multi-mode Cepheids using the database of the ASAS-SN survey has led to the detection of thirteen new double-mode and two triple-mode Cepheids in the Galactic disk. These discoveries have increased the number of Galactic disk multi-mode Cepheids  by 33\%.  One of the new triple-mode variables pulsates simultaneously in the fundamental  and in the first and the second radial overtone modes and the other in the first three radial overtone modes. Overtone triple-mode Cepheids were identified only in the Galactic bulge and in the Large and Small Magellanic Clouds previously.\footnote{During the preparation of this work, Udalski {\it et al.} (2018) announced the discovery of a large number of Cepheids, including numerous double-mode and one new triple-mode Cepheid in the OGLE Galactic disk fields. As the OGLE survey covers only a portion of the sky, besides performing observations with a dynamic range different to that of ASAS-SN, there will be very few common objects between the two samples, when the list of discoveries by Udalski {\it et al.} (2018) becomes available.    }}
{Stars: variables: Cepheids -- Stars: oscillations (including pulsations)}

\Section{Introduction}

Classical Cepheid stars constitute a crucial component of the Cosmic Distance Ladder, as they follow a tight period-luminosity relationship,
while their brightness allows their detection in far-away galaxies (see Catelan and Smith 2015 for a review). These qualities permit the calibration of other distance indicators, contributing immensely to our understanding of the Universe.

While most Classical Cepheids pulsate in one radial mode (usually the fundamental or the first overtone), there are regions within the
instability strip where more than one mode might be excited at the same time. The locations of these regions are heavily affected by the
properties of the stars (Koll\'ath and Buchler 2001, Smolec and Moskalik 2010). For example, the fundamental/first-overtone (F/1O) double mode Cepheids follow a tight relationship between the iron abundance and the periods of the two modes, thus providing a means to estimate photometric metallicities from the pulsation periods alone (see Kovtyukh {\it et al.} 2016 and references therein). Such photometric iron abundances lead to metallicity gradients, which are consistent with spectroscopic values for the Milky Way (Lemasle {\it  et al.} 2018). This technique can be trivially extended to extragalactic Cepheids, which are too faint for individual spectroscopic studies, as was done by  Beaulieu {\it et al.} (2006).

The OGLE Survey (Udalski {\it et al.} 2015) has led to a virtually complete census of multi-mode Cepheids in the Magellanic Clouds (Soszy\'nski {\it  et al.} 2015). Unfortunately, such a census is lacking for the Cepheids in the Milky Way, owing to the large differences in distance, extinction, as well as the large area of the sky over which they could be found.

Here we present the results of a systematic search for multi-mode Cepheids using the wealth of data provided by the ASAS-SN survey\footnote{{\it https://asas-sn.osu.edu/variables}} (Shappee {\it et al.} 2014, Jayasinghe et al. 2018a,b), which led to the discovery of fifteen Galactic multi-mode Cepheids.

\Section{New Double/Triple-Mode Cepheids}

The ASAS-SN $V$-band light curves of variables classified as Classical Cepheids (DCEP and DCEPS classes according to the nomenclature adopted in the General Catalogue of Variable Stars, Kholopov {\it et al.} 1998) with periods up to 5~d,
as well as sources classified as RR~Lyrae variables with periods longer than 0.6~d, were individually inspected, selecting an initial set of stars with light curves resembling those of multi-mode Cepheids. All sources down to 15 mag were considered, totaling 3729 variables from the sample provided by Jayasinghe {\it et al.} (2018a,b). In order not to miss any multi-mode variables with apparently noisy light curves and/or of non-typical shapes, this sample was not cut using the $Prob$ and $LKSL$ parameters as suggested by Jayasinghe {\it et al.} (2018b) to obtain a non-contaminated sample. As a  consequence, many of the variables checked were in fact binaries and non-periodic variables.

\begin{table}[ht]
\begin{scriptsize}\begin{center}
\caption{{\small Identification of new double/triple-mode Cepheids.}\label{table:1}}
\vskip3pt
{\begin{tabular}{r@{\hspace{-2pt}}c@{\hspace{0pt}}l@{\hspace{-1pt}}l@{\hspace{3pt}}r@{\hspace{3pt}}r@{\hspace{5pt}}l} 
 \hline \noalign{\vskip3pt}
\multicolumn{1}{c}{No.  ASAS-SN-V ID} & {Pulsation} & \multicolumn{1}{c}{Period{$^b$}}  & \multicolumn{1}{c}{$\overline{V}$} & \multicolumn{1}{c}{N$^c$}&T$^d$ &Other name\\
 & {modes$^{a}$} & \multicolumn{1}{c}{ [d]}  & \multicolumn{1}{c}{[mag]} & &\multicolumn{1}{c}{[d]}&\\
\noalign{\vskip3pt}
\hline
\noalign{\vskip3pt}
 1\,\,\, J103920.14--545134.7 & {\bf 1O}/2O/3O &  0.600847& 12.648 &350(6)&776 &ASAS J103920-5451.6\\  
 2\,\,\, J060658.07+252402.1  & {\bf{1O}}/2O   &  0.611287& 12.514 &141(1)&1186&DT Gem             \\  
 3\,\,\, J054002.90+160503.1  & {\bf{1O}}/2O   &  0.630811& 14.892 &264(5)&1188&                   \\  
 4\,\,\, J062805.03+142806.6  & {\bf{1O}}/2O   &  0.640708& 14.045 &148(3)&1186&                   \\  
 5\,\,\, J074310.73--113457.7 & {\bf{1O}}/2O   &  0.715356& 12.888 &255(1)&1188&                   \\   
 6\,\,\, J073543.38--313225.8 & {\bf{1O}}/2O   &  0.805315& 14.298 &411(2)& 783&  \\ 
 7\,\,\, J082217.31--461941.1 & {\bf{1O}}/2O   &  0.957680& 14.282& 127(1)& 779&  \\  
 8\,\,\, J065759.86+053444.9  & F/{\bf{1O}}/2O  & 0.978119& 14.580 &167(5)&1186&  \\   
 9\,\,\, J210015.75+482657.7  & {\bf{F}}/1O     & 1.46666 & 12.947 &81(2)& 931& V1543 Cyg\\   
 10\,\,\, J092202.82--570954.3 & {\bf{F}}/1O    & 2.57918 &14.116& 171(3)& 759 & \\   
 11\,\,\, J065046.50--085808.7 & F/{\bf{1O}}    & 2.58786 &14.179& 264(1)&1193&  \\  
 12\,\,\, J205916.93+443346.1  & {\bf{F}}/1O    & 2.79666 &14.532& 247(3)& 975&  \\  
 13\,\,\, J192801.27+195659.6  & {\bf{F}}/1O    & 2.80725 & 14.529& 277(1)& 1133&   \\    
 14\,\,\, J091847.73--500445.3 & {\bf{F}}/1O    & 3.00770 & 14.441& 315(7)& 780& GDS\ J0918478-500445 \\  
 15\,\,\, J192550.01+194925.2  & F/{\bf{1O}}    & 3.50631 & 13.661 &279(2)&1133 &  \\     
\noalign{\vskip3pt}
\hline
\noalign{\vskip3pt}
\multicolumn{7}{l}{$^{a}$ The dominant mode is set in boldface.} \\
\multicolumn{7}{l}{$^{b}$ 1O period determined from the ASAS-SN-V data.}\\
\multicolumn{7}{l}{$^{c}$ The number of data used; the number of removed outliers is given in parenthesis.} \\
\multicolumn{7}{l}{$^{d}$ Time-span of the observations.}
\end{tabular}}
\end{center}
\end{scriptsize}
\end{table}

Further examination of the light curves retrieved from the ASAS-SN Variable Star Database (Shappee {\it et al.} 2014, Jayasinghe 2018a,b) of this initial sample revealed that most of them were either RR~Lyrae stars with the presence of the Blazhko effect, or Classical Cepheids with noisy light curves. Nevertheless, after the removal of the previously known objects, this search has led to the discovery of thirteen new double-mode and two triple-mode Cepheids, increasing the number of such objects in the Galactic field significantly.

Each of the new multi-mode Cepheids  lays at low Galactic latitudes, indicating that they likly belong to the disk population of the Galaxy.

The detailed analysis of the light curves was performed using the program packages MUFRAN (Koll\'ath 1990) and LC{\sc{fit}} (S\'odor 2012).  Outlier points, $i.e.$, data points differing by more than $3\,\sigma$ from the best fit model of the data, were removed in consecutive steps. 

The ASAS-SN identification of the new double/multi-mode Cepheids, the detected radial modes and the period of the first-overtone (1O) mode, which appears in all of the stars, are given in Table~1. The dominant mode is set in boldface.  The mean magnitudes, the number of data points analyzed/removed and the time-span of the observations are also given in Table~1.

The frequencies and amplitudes of the detected radial modes, as well as the frequencies and the S/N ratios of the largest amplitude  signals appearing in the Fourier spectrum of the residuals after the removal of the appropriate number of harmonics for the radial modes and their linear combination terms, are listed in Table~2. The $rms$ scatter of the prewhitened light curves and the mean level of the residual spectra are also given.

\begin{landscape}
   \begin{table}[ht]
\begin{scriptsize}\begin{center}
   \caption{\small {Detected frequencies of the new double/triple-mode Cepheids.}   \label{table:2}}
\vskip3pt
   {\begin{tabular}{rl@{\hspace{4pt}}c@{\hspace{2pt}}l@{\hspace{4pt}}c@{\hspace{2pt}}l@{\hspace{4pt}}c@{\hspace{2pt}}l@{\hspace{4pt}}cccl} 
   \hline \noalign{\vskip3pt}
   {No.} & \multicolumn{2}{c}{F}& \multicolumn{2}{c}{1O} & \multicolumn{2}{c}{2O} & \multicolumn{2}{c}{3O}& \\ 
      & \multicolumn{1}{c}{Freq.$^{a}$} &\multicolumn{1}{c}{ Amp. }  & \multicolumn{1}{c}{Freq.$^{a}$ }&\multicolumn{1}{c}{ Amp.}& \multicolumn{1}{c}{Freq.$^{a}$} &\multicolumn{1}{c}{ Amp.}& \multicolumn{1}{c}{Freq.$^{a}$} &\multicolumn{1}{c}{ Amp. }&$rms^{b}$& Noise$^{c}$& Residual frequencies$^{d}$\\ 
   &\multicolumn{1}{c}{[d$^{-1}$]}&\multicolumn{1}{c}{[mag]} &\multicolumn{1}{c}{[d$^{-1}$]}&\multicolumn{1}{c}{[mag]}&\multicolumn{1}{c}{[d$^{-1}$]}&\multicolumn{1}{c}{[mag]}&\multicolumn{1}{c}{[d$^{-1}$]}&\multicolumn{1}{c}{[mag]}&\multicolumn{1}{c}{[mag]}&\multicolumn{1}{c}{[mag]}&\\
   \noalign{\vskip3pt}
   \hline
   \noalign{\vskip3pt}
   1 & \multicolumn{1}{c}{--} &\multicolumn{1}{c}{--} & 1.664318(4)& 0.197  & 2.07019(3) & 0.028  & 2.46444(3) & 0.024   &0.015 &0.001 & {\it 2.5317(3.7)}, 5.0179(3.4), 5.2270(3.9)\\   
   2  & \multicolumn{1}{c}{--} &\multicolumn{1}{c}{--}& 1.635894(5)  &0.181& 2.03175(2)   &0.034  & \multicolumn{1}{c}{--} & \multicolumn{1}{c}{--} &0.015& 0.002& {\it 1.6350(3.9), 1.9968(4.5), 2.0311(5.6), 3.2590(4.2) }\\    
   3  & \multicolumn{1}{c}{--} &\multicolumn{1}{c}{--}& 1.585262(9)  &0.176& 1.96414(4)   &0.052  & \multicolumn{1}{c}{--} & \multicolumn{1}{c}{--} &0.047& 0.005& 0.6428(3.3), 8.2058(3.8)     \\    
   4  & \multicolumn{1}{c}{--} &\multicolumn{1}{c}{--}& 1.560773(9)  &0.163& 1.93864(5)   &0.025  & \multicolumn{1}{c}{--} & \multicolumn{1}{c}{--} &0.027& 0.004& {\it 1.5600(4.8)}  \\    
   5  & \multicolumn{1}{c}{--} &\multicolumn{1}{c}{--} & 1.397905(8) &0.110 & 1.73186(3) &0.032   & \multicolumn{1}{c}{--} &\multicolumn{1}{c}{--} &0.023& 0.002&2.0019(3.8), 5.8738(3.3)\\           
   6  & \multicolumn{1}{c}{--} &\multicolumn{1}{c}{--} & 1.24175(1)  &0.136 & 1.54034(4)  & 0.038  & \multicolumn{1}{c}{--} & \multicolumn{1}{c}{--} &0.044& 0.004& {\it 1.5414(5.3), 1.5428(3.2)} \\         
   7  & \multicolumn{1}{c}{--} & \multicolumn{1}{c}{--} &1.04419(1)  &0.227&   1.29818(8) &0.024   &\multicolumn{1}{c}{--} & \multicolumn{1}{c}{--} &0.027& 0.004 &  1.8230(3.2), 1.9920(3.2)\\ 
   8  & 0.75065(1)& 0.118    & 1.022371(8)  & 0.157 & 1.27302(6) & 0.015    & \multicolumn{1}{c}{--} & \multicolumn{1}{c}{--} & 0.032& 0.004& 4.6595(4.2), 4.9732(3.5)  \\  
   9  & 0.492536(4) & 0.256  & 0.68182(3)   & 0.051& \multicolumn{1}{c}{--} & \multicolumn{1}{c}{--} &  \multicolumn{1}{c}{--} & \multicolumn{1}{c}{--} & 0.013& 0.002& 8.5497(3.5)    \\     
   10 & 0.27874(1) & 0.171           & 0.38772(1)   & 0.156 & \multicolumn{1}{c}{--} & \multicolumn{1}{c}{--} & \multicolumn{1}{c}{--} & \multicolumn{1}{c}{--} & 0.023& 0.003& 4.2266(3.5) \\   
   11 & 0.277321(7)     &0.139       & 0.386420(6)  & 0.153& \multicolumn{1}{c}{--} & \multicolumn{1}{c}{--} & \multicolumn{1}{c}{--} & \multicolumn{1}{c}{--} & 0.024& 0.003& 2.0028(4.3) \\         
   12 & 0.25312(1)   &   0.146         & 0.35757(2) & 0.075  & \multicolumn{1}{c}{--} & \multicolumn{1}{c}{--} &  \multicolumn{1}{c}{--} & \multicolumn{1}{c}{--} & 0.044& 0.005& {\it 0.3374(3.4)} \\    
   13 & 0.247265(9) &0.211            & 0.35622(2)   & 0.096 & \multicolumn{1}{c}{--} & \multicolumn{1}{c}{--} & \multicolumn{1}{c}{--} & \multicolumn{1}{c}{--} &0.049& 0.005 & {\it 0.2483(3.2)}, 0.9586(3.3) \\
   14 & 0.234005(9)    &0.205        & 0.33248(2)   & 0.092& \multicolumn{1}{c}{--} & \multicolumn{1}{c}{--} & \multicolumn{1}{c}{--} & \multicolumn{1}{c}{--} &0.038& 0.004& 0.8753(3.4), 1.4770(3.4)        \\       
   15 & 0.19860(2)     &0.074        & 0.28520(1)  & 0.145& \multicolumn{1}{c}{--} & \multicolumn{1}{c}{--} & \multicolumn{1}{c}{--} & \multicolumn{1}{c}{--} & 0.038& 0.004&0.0031(3.7)            \\       
   \noalign{\vskip3pt}
   \hline
\noalign{\vskip3pt}
\multicolumn{12}{l}{$^{a}$ 1\,$\sigma$ uncertainty of the last digit is given in parenthesis.} \\
\multicolumn{12}{l}{$^{b}$ $rms$ scatter of the residual after removing the pulsation and the combination frequencies.}\\
\multicolumn{12}{l}{$^{c}$  Mean noise level of the residual spectra.} \\
\multicolumn{12}{l}{$^{d}$  The S/N is given in parentheses. Frequencies close to the radial-mode frequencies and their harmonic components are set in italics. }
\end{tabular}}
\end{center}
\end{scriptsize}   
\end{table}
 \end{landscape}

Although many of the listed residual signals may originate from instrumental/reduction effects, from noise and/or from contamination of neighbouring stars ($e.g.$, frequencies close to integer numbers detected in stars No.\,5 and No.\,11,  frequencies shorter than the fundamental mode frequency, as well as frequencies which are too high), and the S/N ratio of some of them is only $3.2-3.5$,  the first few largest-amplitude residual frequencies are listed for each star for guidance.

In the case of double- and overtone-mode Cepheids, a variety of additional signals appearing in the residual spectra have been detected in the literature -- see $e.g.$, Moskalik and Kolaczkowski (2009), Soszy\'nski {\it et al.} (2015) and Smolec and \'Sniegowska (2016). Additionally, 1O/2O double-mode Cepheids may display anticorrelated modulations of the radial modes, as well as a frequency with $f_{\mathrm{1O}}/f_{\mathrm x}=0.61-0.64$ ratio, which appears quite frequently when the first radial overtone mode is excited.


Significant,  often unresolved, residual signals appear close to one of the detected radial modes in 6 stars in our sample. As the time-base of the ASAS-SN light curves is  $3-4$ years long, these signals may arise from period change. Nevertheless, light-curve variations cannot be excluded as the origin of these residual signals either, although symmetric multiplets at the pulsation components, which are characteristics of modulated light curves, are not detected in any of the residual spectra.  The only star displaying residual signals at two radial modes is star No.\~2., $i.e.$, DT Gem. However, as these signals are not resolved, and other residual signals also appear in this star, higher quality data are needed to unambiguously detect possible correlated modulations (Moskalik and Kolaczkowski 2009) of the pulsation modes.

Some of the detected residual frequencies may correspond to non-radial modes. However, independent observations are needed to establish their true origin. It is important to note, however, that none of the detected residual signals matches the $0.61-0.64$ ratio to the frequency of the 1O mode.

The position of F/1O double-mode Cepheids on the Petersen diagram puts very tight constraints on the metallicity, as it was shown by Buchler and Szab\'o (2007). Formulae to calculate the metallicity of F/1O double-mode Cepheids  were given by Kovtyukh  {\it  et al.} (2016) and Szil\'adi {\it et al.} (2018). 
The period ratios of the detected radial modes and the metallicities of the F/1O variables estimated using formulae based on the periods and the period ratios (columns I, II) and also by using the Fourier amplitudes and phases of the light curves of the corresponding radial modes (columns III and IV) are given in Table 3. 

The overall scatter of the differences between the metallicities given in columns I and II is only 0.04\,dex, which is not surprising taking into account that the method and the calibrating samples of these formulae are mostly identical. The $rms$ scatter of the differences  between the metallicities  derived from the periods (columns I and II) and from the Fourier parameters (columns III and IV) are, however, significantly larger, $0.14-0.18$\,dex, and this is even larger, $0.32$\,dex, for the results derived from the Fourier parameters of the F and the 1O radial modes (columns III and IV). Taking into account the relatively small, $0.5-0.6$ dex overall range of the metallicity values of Galactic Cepheids, these results warn that using the light-curve parameters the derived metallicities of double-mode Cepheids may be unreliable. This is especially true for the overtone mode, whose light curve can be quite sinusoidal, with the end result that  the amplitudes and phases of the harmonic components of this mode can be very uncertain. This is the case for variables No.~11 and No.~15. Although the 1O mode is dominant in these stars, their metallicities calculated from the parameters of this mode are clearly erroneous.

\begin{table}[t]
\begin{scriptsize}
\begin{center}
   \caption{\small {Period ratios and estimated metallicities of the new double/triple-mode Cepheids.}\label{table:3}}
\vskip3pt
   \begin{tabular}{rccccr@{\hspace{15pt}}r@{\hspace{15pt}}r@{\hspace{15pt}}r} 
   \hline \noalign{\vskip3pt}
   {No.} &  \multicolumn{1}{c}{1O/F}& \multicolumn{1}{c}{2O/1O} & \multicolumn{1}{c}{3O/2O} & \multicolumn{1}{c}{3O/1O} &  \multicolumn{4}{c}{[Fe/H]$^{a}$}  \\ 
   &&&&&\multicolumn{1}{c}{I}&\multicolumn{1}{c}{II}&\multicolumn{1}{c}{III}&\multicolumn{1}{c}{IV}\\
   \noalign{\vskip3pt}
   \hline
   \noalign{\vskip3pt}
   1 & \multicolumn{1}{c}{--} & 0.80395                & 0.84003                & 0.67533                &&&&     \\                 
   2 & \multicolumn{1}{c}{--} & 0.80517                & \multicolumn{1}{c}{--} & \multicolumn{1}{c}{--} &&&&     \\                 
   3 & \multicolumn{1}{c}{--} & 0.80710                & \multicolumn{1}{c}{--} & \multicolumn{1}{c}{--} &&&&     \\                 
   4 & \multicolumn{1}{c}{--} & 0.80509                & \multicolumn{1}{c}{--} & \multicolumn{1}{c}{--} &&&&     \\                 
   5 & \multicolumn{1}{c}{--} & 0.80717                & \multicolumn{1}{c}{--} & \multicolumn{1}{c}{--} &&&&      \\                 
   6 & \multicolumn{1}{c}{--} & 0.80615                & \multicolumn{1}{c}{--} & \multicolumn{1}{c}{--} &&&&      \\                 
   7 & \multicolumn{1}{c}{--} & 0.80435                & \multicolumn{1}{c}{--} & \multicolumn{1}{c}{--} &&&&       \\                 
   8 &  0.73423               & 0.80311                & \multicolumn{1}{c}{--} & \multicolumn{1}{c}{--} & $-$0.44 &$-$0.54& $-$0.45& $-$0.05$^{*}$\\                 
   9 &  0.72238               & \multicolumn{1}{c}{--} & \multicolumn{1}{c}{--} & \multicolumn{1}{c}{--} & $-$0.26 &$-$0.33& $-$0.16&    0.13\,\,\\\                 
   10&  0.71892               & \multicolumn{1}{c}{--} & \multicolumn{1}{c}{--} & \multicolumn{1}{c}{--} & $-$0.30 &$-$0.34& $-$0.14& $-$0.16\,\,\\                 
   11&  0.71767               & \multicolumn{1}{c}{--} & \multicolumn{1}{c}{--} & \multicolumn{1}{c}{--} & $-$0.28 &$-$0.31& $-$0.05&    0.98\,\,\\                 
   12&  0.70789               & \multicolumn{1}{c}{--} & \multicolumn{1}{c}{--} & \multicolumn{1}{c}{--} & $-$0.08 &$-$0.09& $-$0.20&    0.22\,\,\\                 
   13&  0.69414               & \multicolumn{1}{c}{--} & \multicolumn{1}{c}{--} & \multicolumn{1}{c}{--} &   0.22  &   0.25&    0.07&    0.21\,\,  \\                 
   14&  0.70382               & \multicolumn{1}{c}{--} & \multicolumn{1}{c}{--} & \multicolumn{1}{c}{--} &   0.00  &   0.00& $-$0.09& $-$0.07\,\, \\                 
   15&  0.69635               & \multicolumn{1}{c}{--} & \multicolumn{1}{c}{--} & \multicolumn{1}{c}{--} &   0.13  &   0.15&    0.24&    0.52\,\, \\                 
\noalign{\vskip3pt}
\hline
\noalign{\vskip3pt}
\multicolumn{9}{l}{$^{a}$ Determined for stars with simultaneous fundamental and first-overtone pulsations using}\\ 
\multicolumn{9}{l}{I)  Eq.~2 of Kovtyukh {\it et al.} (2016);  II) Eq.~2 of Szil\'adi {\it et al.} (2018); III-IV) Eq.~4 of Szil\'adi}\\
\multicolumn{9}{l}{{\it et al.} (2018) for the fundamental and for the 1O mode.} \\
\multicolumn{9}{l}{$^{*}$ We note that the F/1O Cepheid period-metallicity relation have not been tested for triple-}\\
\multicolumn{9}{l}{mode Cepheids yet, therefore this estimate might be significantly biased from the true value.}
\end{tabular}
\end{center}
\end{scriptsize}
\end{table}

\Section{ Triple-mode Cepheids and the Petersen diagram}

\begin{figure}[t]
\includegraphics[width=12.7cm]{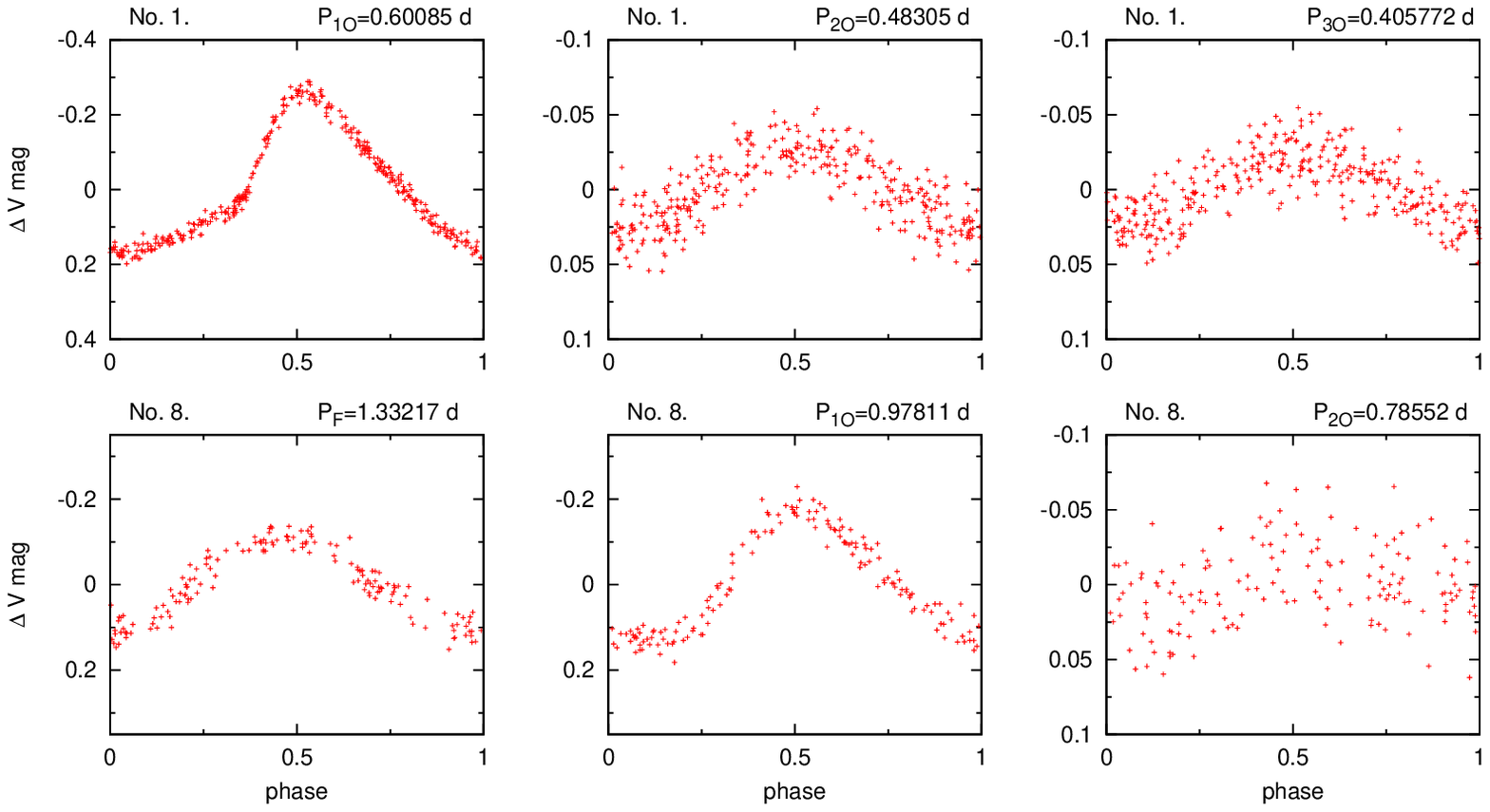}
\vspace*{-3mm}
\FigCap{ASAS-SN light curves folded with the periods of the radial modes of the two new triple-mode Cepheids. Data are prewhitened for the other radial modes and for the linear combination terms. }\label{fig.1}
\end{figure}

Triple-mode Cepheids are very rare objects. Two triple-mode (F/1O/2O) variables (AC And and V823 Cas)  were suspected to belong this class in the Galactic field, however their evolutionary status was ambiguous previously.

The Gaia DR2 measurements (Gaia Collaboration {\it  et al.} 2018) hint at very high luminosities of these stars, indicating that  they are bona fide Cepheids.
The new discovery of a F/1O/2O and a 1O/2O/3O triple-mode Cepheids has increased the number of these stars to four.  
The folded light curves of the individual modes of the new triple-mode variables, prewhitened for the other radial modes and the linear combination terms, are shown in Fig.~1.

There are three 1O/2O/3O Cepheids recognized in the Galactic bulge as well, with extremely short, 0.23--0.30~d 1O periods (Soszy\'nski {\it et al.} 2011,  2017). There are also a few known extragalactic triple-mode Cepheids.  Based on the OGLE-II/III data, the F/1O/2O and 1O/2O/3O modes are simultaneously excited in two  (Soszy\'nski {\it et al.} 2008) and three stars  (Moskalik {\it et al.} 2004, Soszy\'nski {\it et al.} 2008) in the LMC, and in two and one stars  (Soszy\'nski {\it et al.} 2010) in the SMC, respectively. On the other hand, the analysis of the OGLE-IV data of the Magellanic Clouds led to the detection of only one F/1O/2O Cepheid in the LMC and seven and one 1O/2O/3O Cepheids in the LMC and SMC, respectively (Soszy\'nski {\it et al.} 2015).

The period ratio pairs of the new double/triple-mode variables are displayed by filled symbols of different shapes in the Petersen diagram shown in Fig.~2. For comparison, the period ratios of Galactic field and bulge double/triple mode Cepheids are also plotted. 

The F/1O variables compiled by Lamasle {\it et al.} (2018) complemented  with  TYC\,6849\,00019\,1 (Khruslov 2009a) and three F/1O stars of the Galactic bulge (OGLE-BLG-CEP-077, OGLE-BLG-CEP-095, OGLE-BLG-CEP-098, Soszy\'nski {\it  et al.} 2011) are shown by open circles.

\begin{figure}[t]
\includegraphics[width=12.7cm]{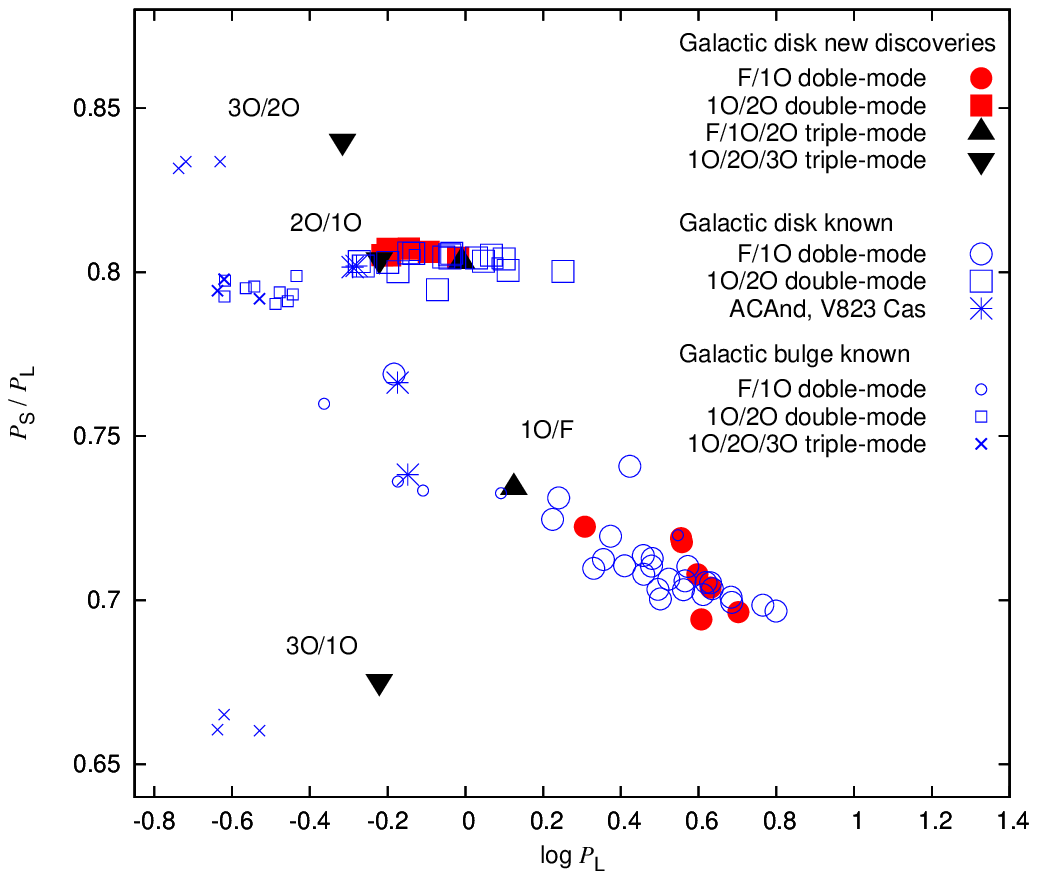}
\vspace*{-3mm}
\FigCap{The period ratios of the new and the known Galactic disk and bulge Cepheids are shown in the Petersen diagram. $P_S$ and $P_L$ stand for the periods of the shorter and longer period mode. The filled and the open symbols denote the new discoveries and known Galactic disk and bulge variables, respectively. The up and down black triangles show the positions of the new F/1O/2O and the 1O/2O/3O Cepheids. The period ratios of the triple mode variables in the Galactic field (AC And and V823 Cas) and in the bulge are indicated by  star and cross symbols, respectively. Galactic bulge stars are shown by small symbols.}\label{fig.2}
\end{figure}

The sample of the 1O/2O Cepheids comprises seventeen field variables: CO Aur (Mantegazza 1983); V363 Cas, V767 Sgr (Hajdu, Jurcsik and S\'odor 2009), V1048 Cen (Beltrame and Poretti 2002); V1837 Aql, V985 Mon, V519 Vel, V356 Mus, V966 Mon, GSC00746-01186, V1345 Cen, V719 Pup, QX Cam (Krushlov 2009b, 2009c, 2010), OGLE-GD-CEP-0004, OGLE-GD-CEP-0006, OGLE-GD-CEP-0014, OGLE-GD-CEP-0018 (Pietrukowicz {\it et al.} 2013), and 10 Galactic bulge variables (Soszy\'nski {\it et al.} 2011). Their period ratios are indicated by open square symbols in Fig.~2. 

The period ratios of known triple-mode field (AC And--Guman 1982; V823 Cas--Antipin 1997) and bulge (3 stars form Soszy\'nski 2011) variables are shown by stars and crosses, respectively.

The 1O/F and 2O/3O period ratios of the new discoveries fit well the period ratios of the known Galactic field variables. The scatter of the 1O/F  ratios is the consequence of  the spread in the metallicities of the stars as documented in Table~3. 

 Taking into account that the non-linear modeling of double-mode pulsation is still controversial (Koll\'ath {\it  et al.} 2002, Szab\'o {\it  et al.} 2004, Buchler 2009, Smolec and Moskalik 2008a,b), the theoretical explanation and modeling of the simultaneous excitation of three radial modes is an extreme challenge for non-linear theory. We do note, however, that at the linear level  there is a consensus of the models. 
E.g., the linear models set strong constraints on the possible evolutionary status and on the parameters of the triple-mode variable, AC And, (Kov\'acs and Buchler 1994) and Moskalik and Dziembowski (2005) succeeded in determining both the physical properties and the evolutionary status of the triple overtone-mode Cepheids in the LMC.

\Acknow{We are grateful to the anonymous referee for the comments, which helped to improve the paper significantly. JJ acknowledges the support of OTKA grant NN-129075. Support for GH and MC is provided the Ministry for the Economy, Development, and Tourism's Millennium Science Initiative through grant IC\,120009, awarded to the Millennium Institute of Astrophysics (MAS); by Proyecto Basal AFB-170002; by Fondecyt through grant \#1171273; and by CONICYT's PCI program through grant DPI20140066. GH also acknowledges funding by the CONICYT-PCHA/Doc\-torado Nacional grant 2014-63140099. The authors would like to thank L. Szabados for providing us a preliminary list of Galactic overtone double-mode Cepheids.
}

\end{document}